%% file: sigconf.tex
\documentclass[sigconf]{acmart}
\AtBeginDocument{%
  }

\setcopyright{none}
\renewcommand\footnotetextcopyrightpermission[1]{}
\copyrightyear{2018}
\acmYear{2018}
\acmDOI{XXXXXXX.XXXXXXX}
\acmConference[Work in progress]{Make sure to enter the correct
  conference title from your rights confirmation email}{June 03--05,
  2018}{Woodstock, NY}
\acmISBN{978-1-4503-XXXX-X/2018/06}

\usepackage{algorithm}
\usepackage{algorithmic}

\begin{document}

\title{AdsWorldEngine: A Self-Evolving Conversational Advertising Agent through Orchestrator and Tool Coevolution}



\author{Simiao Zuo, Chenhui Xu, Yimeng Jia, Qiang Lou, Jian Jiao, Denis Charles}
\affiliation{%
  \institution{Microsoft}
  \city{Redmond}
  \state{WA}
  \country{USA}
}
\email{simiaozuo@microsoft.com}

\renewcommand{\shortauthors}{Zuo et al.}

\begin{abstract}
Conversational advertising aims to deliver useful ads within multi-turn assistant interactions. Unlike conventional query-based advertising, where the user's intent is often expressed in a short standalone query, conversational ads must infer latent commercial intent from the current user query, the assistant response, and dialogue history while also deciding whether an ad would be helpful rather than intrusive. We propose \textbf{AdsWorldEngine}, an agentic framework for conversational advertising. AdsWorldEngine uses an Opportunity Gate to determine whether ads should be shown, an Orchestrator to generate commercial intents, call advertising tools, and construct a top-$3$ ad slate, and an Evaluator to score delivered ads for offline optimization. The central contribution is an iterative actor-tool training procedure: we first train the Orchestrator with supervised fine-tuning and agentic reinforcement learning, then use high- and low-reward rollouts to construct preference data to train tools. This creates a self-improving loop in which the system learns not only how to use advertising tools, but also how to improve them from rewarded behavior. To support subjective production decisions, we introduce label grounded judgment modeling, which trains judgment models from human labels collected under explicit guidelines. It enriches labels with thinking traces, filters inconsistent rationales through reflection, and further optimizes binary judgments with a cost sensitive GRPO variant that preserves asymmetric reward gaps. Offline, AdsWorldEngine improves diversity by $60\%$ and relevance by $80\%$ over the current production ad delivery system. In an online A/B test, it increases RPM by $22\%$ and ads coverage by $74\%$.
\end{abstract}




\maketitle
\fancyhead[LE,RO]{\footnotesize Work in progress}

\input{0-introduction}
\input{0-method}

\input{0-experiments-labeler}
\input{0-experiments-world}
\input{0-experiments-online}
\input{0-related-works}
\input{0-discussions}

\bibliographystyle{ACM-Reference-Format}
\bibliography{sample-base}

\clearpage
\appendix
\input{0-appendix}

\clearpage

\end{document}

%% file: 0-introduction.tex
\section{Introduction}

Online advertising has traditionally matched an advertisement to a compact serving context. In sponsored search, the main signal is the user's query. This query is a short expression of need that supports retrieval, relevance, ranking, and pricing.

Conversational advertising changes this setting. Users now often express needs through interactions with assistants over multiple turns. Their intent may be accumulated, refined, or only implied. A turn such as ``cheaper ones,'' ``anything near me,'' or ``not in the Caribbean'' is not an ad query by itself. It must be interpreted using the earlier turns. The assistant response is also informative. It may surface products, places, or brands that make the user's commercial need more concrete. As a result, ad serving must consider the current user utterance together with dialogue history, optional user context, and the assistant response.

These requirements make conversational advertising a full ad serving problem, not only a retrieval and ranking problem. The system must decide whether the current turn is suitable for ads. It must also infer commercial intents from the conversation, call retrieval and ranking tools, and construct a useful slate. Recent work studies ads in conversational AI, including temporal placement decisions and ad generation that accounts for dialogue~\citep{banchio2024ads,shibata2023conversational}. However, these directions usually address individual components rather than the full ad serving system.

To optimize the advertising experience across the entire interactive process, we introduce \textbf{AdsWorldEngine}, an agentic framework for conversational advertising that is designed for production settings. For each turn, an \emph{Opportunity Gate} decides whether an ad is suitable. Eligible turns pass to an \emph{Orchestrator}. The Orchestrator resolves context, proposes commercial intents, calls advertising tools, and selects a slate with the best ads. An offline \emph{Evaluator} scores the delivered slate from the perspectives of users, advertisers, and publishers. It then supplies rewards for training.

\begin{figure*}[t!]
    \centering
    \includegraphics[width=0.9\linewidth]{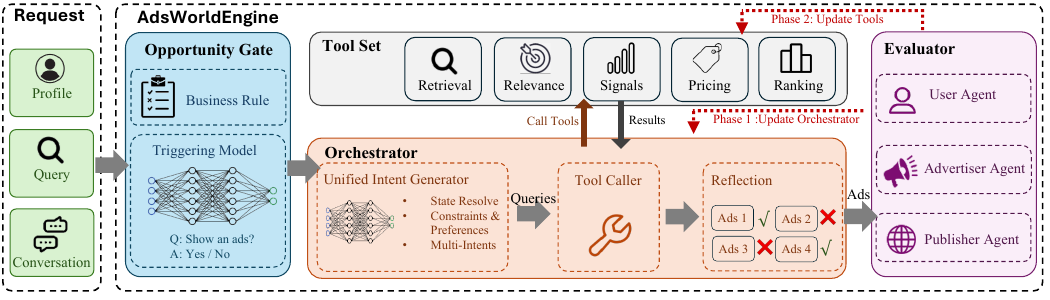}
    \caption{Overview of AdsWorldEngine. At serving time, the system observes the user query, dialogue history, optional user context, and the assistant response. The Opportunity Gate first decides whether ads should be shown. If the turn is eligible, the Orchestrator generates ad-seeking intents, calls advertising tools, and selects a top-$3$ slate. Offline, the Evaluator scores delivered ads and provides rewards for improving both the Orchestrator and the tools.}
    \label{fig:arch}
\end{figure*}

A central technical challenge is optimization. Prompting and supervised fine tuning can teach the Orchestrator to follow a format, generate plausible intents for ads, and call tools correctly. However, they do not fully optimize the interaction between intermediate decisions and downstream tools. For example, two generated ad queries may look diverse to humans but retrieve the same merchant or landing page. Another pair may look similar but retrieve complementary ads because of inventory and retrieval model behavior. We therefore propose \emph{iterative actor and tool optimization}. The Orchestrator is first trained with supervised fine tuning and agentic reinforcement learning using Evaluator rewards. Then, rollouts with high and low rewards are converted into preference data for retrieval, relevance, and ranking tools. This creates a loop that improves over time. The Orchestrator learns how to use the tools, and its rewarded behavior produces data that improves the tools themselves.

A second challenge is judgment modeling under production standards. Many reward modeling approaches rely on LLM judge prompts or rubrics fitted by prompts~\citep{liu2023geval,zheng2023judging}. In conversational advertising, however, judgments are not purely verifiable. They depend on usefulness, relevance, and user experience. Their errors also have asymmetric costs across tasks. For example, an inappropriate ad exposure can harm user trust more than a missed opportunity. Selecting an irrelevant ad can also be worse than leaving a slot empty. We therefore propose \emph{label grounded judgment modeling}, a data generation pipeline for training production judgment models. The pipeline starts with task specific production guidelines and human labels collected under those guidelines. For each labeled example, we generate a thinking trace conditioned on the label. The trace explains how the assigned label aligns with the guideline and describes which types of recommendations may be helpful without being intrusive to the user-assistant interaction. This differs from asking the model to invent its own judgment. A reflection step then checks whether the trace supports the label, follows the guideline, and avoids unsupported assumptions. Examples with inconsistent rationales are removed. The remaining pairs of labels and traces provide dense supervision for judges. To optimize decisions with asymmetric costs, we further introduce a GRPO variant with cost sensitive rewards~\citep{shao2024deepseekmath}. This variant uses asymmetric rewards and omits group standard deviation scaling, which preserves task specific penalty gaps. The main contributions of this paper are:
\begin{itemize}
    \item We formulate conversational ad serving as a \textbf{gated agentic framework for production settings}. The framework uses turn context and assistant responses to decide when to show ads and how to construct slates with the best ads.

    \item We propose \textbf{iterative actor and tool optimization}. In this approach, rewarded Orchestrator rollouts train both the actor and the retrieval, relevance, and ranking tools. This enables the system to learn how to call tools and how to improve them from rewarded behavior.

    \item We introduce \textbf{label grounded judgment modeling with cost sensitive GRPO}. For data, the method generates label and trace pairs from task guidelines and filters them through reflection. For training, it uses cost sensitive GRPO with asymmetric rewards for production judgment tasks.

    \item AdsWorldEngine improves diversity by 60\% and relevance by 80\% over the current production ad serving system. It also increases RPM (Revenue Per Mille) by 22\% and ads coverage by 74\% in an online A/B test.
\end{itemize}

%% file: 0-method.tex
\section{AdsWorldEngine Overview}

\label{sec:method_overview}

AdsWorldEngine formulates conversational advertising as a gated agentic decision process. At each turn, it observes the user query, dialogue history, optional user context allowed by policy, and the assistant response, whose recommendations or context carried over from earlier turns can clarify intent. As Fig.~\ref{fig:arch} illustrates, the online pipeline separates \emph{whether} to show ads from \emph{which} ads to show. 

Our framework primarily consists of four components: \emph{(1) Opportunity Gate, (2) Orchestrator, (3) Tool Set}, and \emph{(4) Evaluator}.
The \emph{Opportunity Gate} combines production rules with a learned triggering model to abstain or pass a suitable turn. The \emph{Orchestrator} then resolves the conversation state, generates commercial intents, calls advertising tools from \emph{Tool Set}, and selects a slate with three ads. This adapts language model agents that use tools to ad serving, where dialogue reasoning must be combined with retrieval, ranking, relevance, pricing, and other advertising signals~\citep{yao2023react,schick2023toolformer}. Finally, an \emph{Evaluator} provide feedback signals to update the \emph{Orchestor} and \emph{Tool Set}. We introduce these compontents in details in Sec.\ref{sec:core_components}.

For subjective decisions such as ad triggering and conversation-to-ad relevance, we introduce \emph{label grounded judgment modeling}, as will be discussed in Sec. \ref{sec:label_grounded_judgment}. The method first generates thinking traces conditioned on guidelines and human labels, then filters them through a reflection step to remove inconsistent reasoning.
For asymmetric binary tasks, we use cost sensitive GRPO, which preserves raw reward gaps and allows false positive triggers to carry a larger penalty than false negatives.

We also propose iterative actor and tool optimization. The Orchestrator is trained with supervised fine tuning (SFT) and agentic reinforcement learning from Evaluator rewards. Comparisons between rollouts with high and low rewards then provide preference data for retrieval, relevance, and ranking tools. Alternating these updates lets rewarded Orchestrator behavior improve the tools used in later rollouts. We introduce details in Sec.\ref{sec:actor_tool_iterative_training}.

\section{AdsWorldEngine Core Components}
\label{sec:core_components}
\subsection{Opportunity Gate}
\label{sec:opportunity_gate}

The Opportunity Gate is the first online stage of AdsWorldEngine. Given the full turn context, it decides whether ads should be shown. If the gate rejects the turn, the system abstains. If it accepts the turn, the conversation is passed to the Orchestrator for intent generation, tool calling, and slate selection.

The gate combines production rules with a learned triggering model. The rules enforce hard constraints such as policy and privacy requirements. The learned model decides whether an ad would help the current task. The gate triggers for purchase planning or concrete options surfaced by the assistant response. It abstains in sensitive, purely informational, or disruptive contexts. Overall, the gate favors user experience and treats false positives as especially costly.

The triggering model is trained with the label grounded judgment modeling framework in Section~\ref{sec:label_grounded_judgment}. Human labels are collected under the triggering guideline. They are then converted into label grounded thinking traces, filtered by reflection, and used for SFT and cost sensitive GRPO. This produces a judgment model that separates the production decision of \emph{whether} to show ads from the downstream decision of \emph{which} ads to show. 

\begin{figure*}[t]
\centering
\fbox{
\begin{minipage}{0.95\linewidth}
\small
\textbf{Input conversation.}
The user first asks for beach holidays above $28^\circ$C at Christmas, with direct flights from Heathrow and flight time no longer than 12 hours. After seeing options that include Caribbean destinations, the user asks: ``Anything not in the Caribbean.''

\vspace{0.4em}
\textbf{State resolution.}
The current turn is a refinement, not a topic shift. The omitted target is the earlier request for warm Christmas beach holidays with direct Heathrow flights under 12 hours. The new constraint is to exclude Caribbean destinations.

\vspace{0.4em}
\textbf{Constraint extraction.}
Hard constraints: beach destination; Christmas timing; typical temperature above $28^\circ$C; direct flight from Heathrow; flight time $\leq$12 hours; not in the Caribbean. Assistant-suggested attributes such as luxury or family resorts are not user constraints.

\vspace{0.4em}
\textbf{Intent and query generation.}
[\textit{Indian Ocean beach holidays Christmas direct from London}, \textit{Middle East beach holidays Christmas direct from London}, \textit{Africa beach holidays Christmas direct from London}]

\vspace{0.4em}
\textbf{Tool results.}
The retrieval/relevance/ranking tools return specific Egypt hotel pages, broad Africa beach-holiday pages, repeated Asilia Africa safari-and-beach pages, repeated Ras al Hikmah pages, and one UK inland accommodation listing.

\vspace{0.4em}
\textbf{Reflection and final selected ads.}
Reflection selects \texttt{a7} and \texttt{a9} as strong specific matches, keeps \texttt{a1} for broader Africa coverage, removes near-duplicates, and rejects the UK property as off-intent. The final selected ads are \texttt{a7, a9, a1}.
\end{minipage}
}
\vskip -0.1in
\caption{Example Orchestrator trajectory. Intermediate traces teach the model to resolve a short follow-up, preserve inherited constraints, generate useful retrieval queries, and recover from noisy or duplicate tool results.}
\label{fig:orchestrator_example}
\end{figure*}

\subsection{Orchestrator}
\label{sec:orchestrator_training}

The Orchestrator is the actor in AdsWorldEngine. Once the Opportunity Gate accepts a turn, the Orchestrator constructs the final slate with three ads. This is not simply ranking a fixed candidate set. The model must resolve the user's current intent and preserve constraints from earlier turns. It must also generate intents (ad queries) for retrieval, call tools, inspect retrieved candidates, and select a slate that is relevant and diverse.

The Orchestrator has three components. The \textit{intent generator} resolves dialogue state, extracts constraints and preferences, and produces one to three intents. The \textit{tool caller} submits these intents to advertising tools. These tools include retrieval, relevance and ranking tools. They can also include pricing, and signal models. The \textit{reflection} module reviews the returned candidates and selects exactly three ads for the final slate.

\subsubsection{Data Generation}
\label{sec:orchestrator_data_generation}

We generate supervised trajectories with intermediate traces before the final ad selection. Each trajectory teaches a reusable decision process. It covers state resolution and constraint extraction. It also covers intent generation, tool calling, and reflection. This is especially important for conversational follow up turns, where the current user query may be underspecified without earlier turns.

Figure~\ref{fig:orchestrator_example} illustrates an example of the Orchestrator's thinking trace. The model learns that the short current turn is a refinement of the earlier request, not a standalone query. It also learns to preserve user constraints and avoid adopting unsupported assistant suggestions as user requirements. The trace further teaches the model to generate grounded retrieval queries and handle imperfect tool outputs.

\subsubsection{Model Training}
\label{sec:orchestrator_model_training}

We train the Orchestrator in two stages. First, supervised fine tuning teaches the model the basic procedure for tool use. The model learns to resolve the conversation, extract constraints, generate intents, call tools, inspect candidates, and output three ads. SFT is useful for format control and valid tool calls. It also helps with constraint preservation and decomposition.

Second, we apply agentic reinforcement learning as will discuss in Sec.\ref{sec:actor_tool_iterative_training}. The Orchestrator samples complete trajectories that include intents, tool calls, reflection, and final slate selection. The Evaluator (see Section~\ref{sec:evaluator}) scores the final slate using system level objectives such as relevance and diversity. We then use GRPO to reinforce trajectories that produce better final slates~\citep{schulman2017ppo,shao2024deepseekmath}.

The main advantage of agentic reinforcement learning is that it aligns intermediate tool use with the final serving objective. For example, diversity is not rewarded merely because the Orchestrator writes intents that look different. It is rewarded only when those intents retrieve complementary ads that improve the final slate. Thus, GRPO turns instructions such as ``generate diverse intents'' from prompt level preferences into behaviors grounded in outcomes. The Orchestrator learns which intent decompositions, tool calls, and reflection decisions produce better ads for the user.

\subsection{Evaluator}
\label{sec:evaluator}

The Evaluator is an offline component for measuring ad quality and producing training signals. Its scores are not used directly in online serving. Instead, they provide rewards for Orchestrator training and preference data for improving downstream tools. We use the term broadly to cover user, advertiser, and publisher perspectives. The user perspective measures whether the slate is useful to the user. It focuses on relevance and diversity. The advertiser and publisher perspectives capture production requirements.

In the current implementation, we focus on two user perspective signals: relevance for each ad and diversity for the slate. The main learned evaluator is a conversation-to-ad relevance model. This model judges whether a candidate ad helps the user's current task given the conversation context. We train it with the label grounded judgment modeling framework in Section~\ref{sec:label_grounded_judgment}. Human labels are collected under a relevance guideline, enriched with thinking traces, filtered by reflection, and optimized with SFT and GRPO.

Relevance is evaluated at the individual ad level, but a useful slate should also avoid redundancy. Because only three ads are shown, repeated products or pages can noticeably reduce utility. We therefore measure slate level diversity with pairwise cosine similarity among the selected ad embeddings. We penalize slates that contain near duplicate ads.

\section{Label Grounded Judgment Modeling}
\label{sec:label_grounded_judgment}

Many production decisions in conversational advertising are subjective rather than directly verifiable. For example, deciding whether to show ads depends on whether an ad would be commercially useful and whether it would fit the user's experience. It also depends on sensitivity and product policy. Existing LLM judge methods often prompt a model to approximate human preferences~\citep{liu2023geval,zheng2023judging}. RLHF methods instead train reward models from human feedback~\citep{ouyang2022training}. In AdsWorldEngine, we take a label grounded approach. We first define explicit production guidelines and collect human labels under those guidelines. We then train judgment models to match the resulting preferences.

This framework is designed for production settings where errors often have asymmetric costs. For example, in ad triggering, a false positive can be more costly than a false negative. Showing an ad in an inappropriate context may harm user trust, while missing one commercial opportunity is usually less visible. Similar asymmetry appears in relevance modeling. Selecting an irrelevant ad for the final slate can be worse than failing to retrieve one relevant ad.

\subsection{Data: Label Grounded Thinking Traces}
\label{sec:lgjm_data}

The data pipeline has four steps. First, human annotators assign labels using a task specific guideline. Second, for each labeled example, we generate a thinking trace conditioned on the guideline, the input, and the human label. Third, we use a reflection prompt to check whether the trace supports the label and follows the guideline. Finally, only examples that pass reflection are used for supervised fine tuning.

The trace generation prompt asks the model to explain why the human label is correct under the guideline. This converts a categorical label into denser supervision. The model learns the target judgment and the decision pattern expected by the production policy. However, generated traces may contradict the label, add unsupported assumptions, or fail to follow the required format. We therefore filter examples by checking whether
\[
\text{input} \rightarrow \text{thinking trace} \rightarrow \text{label}
\]
is internally consistent.
Reflection filtering is important because inconsistent rationales teach the model to produce plausible explanations that do not support the target decision. Keeping only traces that follow the guideline improves the reliability of judgments. See Table~\ref{tab:reflection_filtering} for an example.

\begin{table}[t]
\centering
\small
\caption{Example of data filtering. The trace is fluent but contradicts the human label, so it is removed before SFT.}
\label{tab:reflection_filtering}
\vskip -0.1in
\begin{tabular}{p{0.24\linewidth} p{0.67\linewidth}}
\toprule
\textbf{Field} & \textbf{Content} \\
\midrule
Conversation &
User asks for a kid-friendly camping resort near Newton, MA with cabins, AC, kitchen, and teen activities. The assistant recommends specific campgrounds and links. \\
\midrule
Human label &
\texttt{No} (information-seeking conversation) \\
\midrule
Generated trace &
``... directly about commercial entities ... ads for camping resorts or booking platforms would be relevant and helpful ... the human label of \texttt{No} is incorrect ...'' \\
\midrule
Reflection result &
\texttt{REASONING\_LABEL\_MISMATCH}. The trace supports the opposite label and questions the given label. \\
\bottomrule
\end{tabular}
\end{table}

\subsection{Training: SFT and Cost Sensitive GRPO}
\label{sec:lgjm_training}
After filtering, we train judgment models in two stages. Supervised fine tuning teaches the model to follow the task format and produce reasoning grounded in the guideline. We then apply GRPO~\citep{shao2024deepseekmath} to optimize task rewards. These rewards capture label correctness, format validity, and guideline consistency.

For asymmetric binary judgment tasks, we use cost sensitive rewards. For example, in ad triggering, let \texttt{Yes} mean that ads should be shown and \texttt{No} mean that ads should not be shown. We use
\[
R(y,\hat{y}) =
\begin{array}{c|cc}
 & \hat{y}=\texttt{Yes} & \hat{y}=\texttt{No} \\
\hline
y=\texttt{Yes} & 1 & -1 \\
y=\texttt{No} & -2 & 1
\end{array}
\]
This reward directly encodes the production preference that false positive ad triggering should be penalized more strongly than false negatives.

Standard GRPO computes group relative advantages by centering and scaling rewards within a group:
\[
A_i^{\mathrm{scaled}}
=
\frac{R_i-\bar{R}}{s_R+\epsilon},
\]
where $\bar{R}$ and $s_R$ are the group mean and standard deviation. This scaling can be undesirable for binary tasks because the raw reward gap encodes the cost of different mistakes. Dividing by the group standard deviation can remove that information.

We therefore use group centering without group standard deviation scaling:
\[
A_i^{\mathrm{none}} = R_i-\bar{R}.
\]
This preserves the within group comparison while keeping the magnitude of the reward gap. The motivation is related to recent analyses of normalization in GRPO training~\citep{liu2025understanding}. Our setting, however, is specifically cost sensitive binary judgment.

The effect can be seen in a group with two possible rewards. Suppose completions receive either a better reward $H$ or a worse reward $L$, and $k$ out of $G$ completions receive $H$. Then
\[
\bar{R}=\frac{kH+(G-k)L}{G}.
\]
Without standard deviation scaling,
\[
A_H^{\mathrm{none}}
=
\frac{G-k}{G}(H-L),
\qquad
A_L^{\mathrm{none}}
=
-\frac{k}{G}(H-L).
\]
Thus, the learning signal is proportional to the raw reward gap $H-L$. With group scaling,
\[
s_R = \frac{\sqrt{k(G-k)}}{G}|H-L|,
\]
so
\[
A_H^{\mathrm{scaled}}
=
\sqrt{\frac{G-k}{k}},
\qquad
A_L^{\mathrm{scaled}}
=
-\sqrt{\frac{k}{G-k}}.
\]
The gap $H-L$ cancels out. In other words, scaling preserves which completion is better, but loses how much better it is.

We use this cost sensitive GRPO variant for binary judgment models such as the Opportunity Gate. The same label grounded framework is also used for conversation-to-ad relevance modeling. Each task uses its own labels, guidelines, traces, and rewards. We also include a disscusion of serving distribution shift in Appendix~\ref{apx:appendix}.

\section{Orchestrator and Tool Iterative Training}
\label{sec:actor_tool_iterative_training}

AdsWorldEngine is optimized with an iterative training loop for the actor and tools. The Orchestrator is the actor. It resolves the conversation, generates intents, calls tools, and selects the final slate. The tools define the actor's action space. They include retrieval and relevance tools, along with pricing, signal, and ranking models. The key contribution is that we do not treat these tools as fixed infrastructure. Instead, we use rewarded Orchestrator rollouts to improve both the actor and the tools.

\begin{algorithm}[t]
\caption{Iterative optimization for actor and tools.}
\label{alg:actor_tool}
\begin{algorithmic}[1]
\REQUIRE Conversations $\mathcal{D}$, initial tools $\mathcal{T}_0$, Evaluator $E$, Orchestrator SFT data $\mathcal{D}_{\mathrm{sft}}$
\STATE Train initial Orchestrator $\pi_{\theta_0}$ on $\mathcal{D}_{\mathrm{sft}}$
\FOR{$t = 0, 1, \ldots, T-1$}
    \STATE \textbf{Orchestrator update with fixed tools}
    \STATE Sample rollouts from $\pi_{\theta_t}$ using tools $\mathcal{T}_t$
    \STATE Score final slates with Evaluator $E$
    \STATE Update $\pi_{\theta_t} \rightarrow \pi_{\theta_{t+1}}$ with GRPO
    \STATE \textbf{Tool preference construction}
    \STATE Generate rollouts from $\pi_{\theta_{t+1}}$ on $\mathcal{D}$
    \STATE Compare high-reward and low-reward rollouts for the same conversation
    \STATE Construct tool preference data $\mathcal{P}_t=\{(\text{intent},a^+,a^-)\}$
    \STATE \textbf{Tool update}
    \STATE Optimize $\mathcal{T}_t \rightarrow \mathcal{T}_{t+1}$ using $\mathcal{P}_t$ with DPO or ranking losses
\ENDFOR
\RETURN Final Orchestrator $\pi_{\theta_T}$ and tools $\mathcal{T}_T$
\end{algorithmic}
\end{algorithm}

Algorithm~\ref{alg:actor_tool} summarizes the procedure. We first train a stable initial Orchestrator with SFT. Each iteration then alternates between two updates. First, with tools fixed, we optimize the Orchestrator using GRPO and Evaluator rewards. This improves intent decomposition, tool use strategy, and reflection behavior under the current tool environment. Second, we roll out the improved Orchestrator to construct preference data for the tools. For a conversation, an ad from a slate with a high reward becomes a positive example. An ad retrieved or selected in a slate with a low reward becomes a hard negative. To reduce noise, we construct such pairs only when the reward gap is sufficiently large.

This loop converts final rewards at the slate level into supervision for intermediate tools. For retrieval, rollouts with high rewards identify ads that should be retrievable for an intent. Rollouts with low rewards expose candidates that were retrieved but should not be preferred. For relevance and ranking, the same rollouts provide realistic preference pairs discovered by the Orchestrator.

The resulting system improves along two coupled dimensions. GRPO makes the Orchestrator better at using the current tools. At the same time, preference learning from rollouts makes the tools better for future Orchestrator decisions. Better tools expand the actor's effective action space. A stronger actor then produces more informative rollouts for the next tool update. This feedback loop between the actor and tools is the central mechanism of AdsWorldEngine. The system learns how to call tools and how to improve them from its own rewarded behavior. This connects conversational ad serving to recent work on optimizing compound language model systems and systems augmented with retrieval~\citep{khattab2024dspy,li2025ragddr}.

\paragraph{Example: preference training for intent-to-ads relevance models.}
For each conversational intent \(q\), we construct a preference triple
\((q,a^{+},a^{-})\), where \(a^{+}\) is an ad from a high-reward slate and
\(a^{-}\) is a hard negative from a low-reward slate. Let \(s_{\theta}(q,a)\) be the score from the trainable relevance model
and \(s_{\mathrm{ref}}(q,a)\) the score from a frozen reference model.
We optimize
\begin{equation}
\begin{aligned}
\mathcal{L}
={}&-\mathbb{E}_{(q,a^{+},a^{-})}
\log \sigma\Big(
\beta \big[
s_{\theta}(q,a^{+})-s_{\theta}(q,a^{-})
\\
&\qquad\qquad
-s_{\mathrm{ref}}(q,a^{+})
+s_{\mathrm{ref}}(q,a^{-})
\big]
\Big).
\end{aligned}
\end{equation}
For an encoder, \(s_{\theta}\) can be the relevance logit. For a
one-token decoder, we define
$
s_{\theta}(q,a)
=\log p_{\theta}(\texttt{Yes}\mid q,a)
-\log p_{\theta}(\texttt{No}\mid q,a).
$

The proposed algorithm is similar to DPO \citep{rafailov2023direct} because it directly increases
the preference margin of \(a^{+}\) over \(a^{-}\), relative to a frozen
reference model. In particular, the scores induce a policy over the
candidate pair \(C=\{a^{+},a^{-}\}\):
\begin{equation*}
\begin{aligned}
\pi_{\theta}(a\mid q,C)
={}&
\frac{\exp s_{\theta}(q,a)}
{\sum_{a'\in C}\exp s_{\theta}(q,a')},
\\
\log
\frac{\pi_{\theta}(a^{+}\mid q,C)}
{\pi_{\theta}(a^{-}\mid q,C)}
={}&
s_{\theta}(q,a^{+})-s_{\theta}(q,a^{-}).
\end{aligned}
\end{equation*}
Therefore, the objective applies the DPO preference principle to relevance
scores rather than autoregressive sequence probabilities.

%% file: 0-experiments-labeler.tex
\begin{table*}[t]
\centering
\small
\caption{Example of Opportunity Gate. The prompted GPT-5 judge and AdsWorldEngine (SFT only) overreact to product language and produce a false positive. AdsWorldEngine recognizes informational intent and correctly suppresses the trigger.}
\label{tab:triggering_example}
\begin{tabular}{p{0.14\linewidth} p{0.77\linewidth}}
\toprule
\textbf{Field} & \textbf{Content} \\
\midrule

Conversation &
\textbf{User:} ``Which shade fits me?''

\textbf{Bot:} ``Choosing the right concealer shade depends on your
skin undertone, depth, and the purpose \ldots\ Step 1: Identify Your Undertone \ldots\ Step 2: Match Depth
\ldots\ Step 3: Purpose of Use \ldots\ To narrow it down, could you tell me:\ldots'' \\

\midrule

Human label &
\texttt{No}. The user is seeking personalized shade-matching information, not
a retailer or purchase option. The assistant can complete the task directly,
so inserting an ad would distract from the current interaction. \\

\midrule

GPT-5 prompt &
``The user is seeking help choosing a concealer shade, which is a beauty
product \textbf{they would likely need to purchase}. The bot is guiding them
toward identifying the right shade, which could \textbf{directly lead to a
buying decision}. Ads for concealers or beauty products would be relevant and
helpful here \ldots''

\textbf{Prediction:} \texttt{Yes} (incorrect). \\

\midrule

AdsWorldEngine \ \ (SFT only) &
``The user's message is clearly about finding the right concealer shade
\ldots\ the conversation is \textbf{directly about selecting a product
variant}. The user's goal is to identify which specific shade to choose, which
is a \textbf{step toward a purchase decision}. An ad could support their
journey rather than distract \ldots''

\textbf{Prediction:} \texttt{Yes} (incorrect). \\

\midrule

AdsWorldEngine &
``The assistant is giving general advice based on the user's undertone, skin
depth, and existing swatches. The conversation is \textbf{about matching a
shade to the user's skin, not about finding where to buy concealer}. The focus
is on \textbf{completing a task that the assistant can handle directly}, so
inserting ads would distract from the user's intent.''

\textbf{Prediction:} \texttt{No} (correct). \\

\bottomrule
\end{tabular}
\end{table*}

\section{Experiments: Label Grounded Judgment}

\subsection{Opportunity Gate: Triggering Models}
\label{sec:triggering_experiments}

Triggering is not merely detecting product mentions. It is a production judgment: would an ad advance the user's current goal without harming the conversation? We use label grounded judgment modeling to tie predictions to production requirements. Because a bad trigger is visible to the user, the target operating point emphasizes fewer false positives while preserving valid opportunities.

We wrote annotation guidelines for Opportunity Gate that specify when an ad is useful and when it should be suppressed because the assistant can answer directly. With these guidelines, we collected \(30{,}000\) human labeled conversations. For each conversation, we prompted GPT-5 with the guideline and human label to produce a reasoning trace, then used a reflection prompt to remove traces that contradicted the guideline or label, leaving \(27{,}000\) examples. Starting from Qwen3-30B-A3B-Thinking, we used \(24{,}000\) examples for SFT and \(3{,}000\) for GRPO.

\paragraph{Results.}
Table~\ref{tab:triggering_results} reports relative changes from the GPT-5 prompt baseline. AdsWorldEngine with SFT substantially reduces FPR, but moves the decision boundary too far: TPR decreases by \(17.19\%\), resulting in a \(4.09\%\) reduction in balanced accuracy. Thus, SFT yields a conservative model that misses the desired production operating point.
In contrast, AdsWorldEngine with SFT and cost-sensitive GRPO preserves baseline TPR while improving FPR by \(39.07\%\), yielding a \(2.51\%\) balanced accuracy gain. The resulting 30B model outperforms the GPT-5 prompted judge while better avoiding unnecessary triggers, and runs much faster than the GPT-5 prompt for a latency sensitive online gate.

\begin{table}[t]
    \centering
    \small
    \caption{Relative changes in True Positive Rate (TPR), False Positive Rate (FPR), and Balanced Accuracy.}
    \label{tab:triggering_results}
    \begin{tabular}{l|ccc}
        \toprule
        Model & \(\Delta\)TPR ($\uparrow$) & \(\Delta\)FPR ($\downarrow$) &
        \(\Delta\)Balanced Acc. ($\uparrow$) \\
        \midrule
        GPT-5 prompt & -- & -- & -- \\
        AdsWorldEngine (SFT) & \(-17.19\%\) & \(\mathbf{-75.42\%}\) & \(-4.09\%\) \\
        AdsWorldEngine & \(\mathbf{0.00\%}\) &
        \(-39.07\%\) & \(\mathbf{+2.51\%}\) \\
        \bottomrule
    \end{tabular}
\end{table}

These results show why reinforcement learning is preferable to prompt engineering or SFT alone. A prompt can state the policy, but it cannot precisely set the model's implicit preference for positive or negative predictions. SFT transfers labeled reasoning patterns, but its token level objective does not optimize the final tradeoff between false positives and false negatives. GRPO directly adjusts the decision boundary for the label grounded production objective, reducing false positives without the large recall loss seen with SFT.

\paragraph{Example.}
Table~\ref{tab:triggering_example} shows a representative case. The user asks which concealer shade fits them, and the assistant responds with guidance about undertone, skin depth, and existing swatches. The correct decision is \texttt{No}: the user wants information, not a retailer or purchase path, so an ad would distract.
This example highlights a bias that prompts and SFT struggle to remove: product related words can produce a trigger even when the goal is noncommercial. Production aligned reinforcement learning teaches that mentioning or selecting a product is not enough evidence that an ad would improve the experience.

\subsection{Evaluator: Conversation-Ads Relevance}
\label{sec:relevance_experiments}

The conversation to ads relevance model follows the same motivation and training procedure as the
triggering model. Its purpose is not only to identify ads that are related to
the conversation, but also to match the production preference for preserving
user experience. In particular, a false positive allows an irrelevant ad to
enter the final slate and can be more disruptive than filtering out one
potentially relevant ad. We therefore optimize the model toward fewer false
positive predictions while retaining useful ads.

We first defined a human annotation guideline for conversation to ads relevance
in Evaluator and collected \(32{,}000\) labeled examples. For each conversation
and candidate ad, we generated a reasoning trace conditioned on the guideline
and human label. A reflection prompt then removed traces that were inconsistent
with the intended judgment, leaving \(30{,}000\) examples. We used \(27{,}000\)
examples for SFT and \(3{,}000\) examples for GRPO, starting from
Qwen3-30B-A3B-Thinking.

Table~\ref{tab:relevance_results} reports changes relative to the GPT-5 prompt. SFT improves TPR, but it also produces more false positives,
suggesting that SFT alone favors accepting ads too broadly.
GRPO directly adjusts this preference toward the desired production operating
point. It reduces FPR by \(12.71\%\) with only a \(0.78\%\) decrease in TPR,
resulting in a \(7.51\%\) improvement in balanced accuracy. This tradeoff is
better aligned with the user experience objective than either prompt based
judgment or SFT alone.

\begin{table}[h]
\centering \small
\caption{Relative changes in True Positive Rate (TPR), False Positive Rate (FPR), and Balanced Accuracy.}
\label{tab:relevance_results}
\begin{tabular}{l|ccc}
\toprule
Model & \(\Delta\)TPR ($\uparrow$) & \(\Delta\)FPR ($\downarrow$) & \(\Delta\)Balanced Acc. ($\uparrow$) \\
\midrule
GPT-5 prompt & -- & -- & -- \\
AdsWorldEngine (SFT) &
\(+10.04\%\) &
\(+6.36\%\) &
\(+3.39\%\) \\
AdsWorldEngine &
\(-0.78\%\) &
\(\mathbf{-12.71\%}\) &
\(\mathbf{+7.51\%}\) \\
\bottomrule
\end{tabular}
\end{table}

%% file: 0-experiments-world.tex
\section{Experiments: Orchestrator and Tools}

\subsection{Training and Evaluation}
\label{sec:orchestrator_tool_training_evaluation}

We evaluate the iterative procedure in Section~\ref{sec:actor_tool_iterative_training}, using Qwen3-30B-A3B-Thinking as the base model for the Orchestrator. The full pipeline begins with supervised warmup and then alternates Orchestrator training and tool training for three rounds. In each round, the Orchestrator is first updated while the tool set is fixed. The improved Orchestrator is then used to generate preference data for updating the intent-to-ads relevance model in the tool set, which is deployed in the next round.

\paragraph{Evaluator and metrics.}
We use a fixed Evaluator throughout both training and final evaluation. Its conversation-to-ads relevance model observes the complete dialogue and judges whether each selected ad is relevant. The Evaluator also measures slate diversity from the three pairwise cosine similarities among ad embeddings, counting a pair as diverse when its similarity is below $0.3$. During training, these relevance and diversity judgments generate reward signals for Orchestrator optimization, and the relevance judgments additionally provide supervision for tool training. For final evaluation, all systems are run on the same conversation set and return exactly three ads per conversation. We report the Evaluator's relevance score, defined as the number of selected ads judged relevant, and its diversity score, defined as the number of diverse ad pairs. The current Production System is the baseline, and all results are reported as relative gains rather than absolute metric values.

\paragraph{Orchestrator warmup.}
We first train the Orchestrator with supervised reasoning traces generated by prompting. Each trace covers state resolution, constraint extraction, intent and query generation, tool interaction, and reflection over retrieved candidates. This stage teaches the model a stable problem solving pattern. In particular, the Orchestrator learns to resolve short follow up turns, preserve constraints from earlier messages, form valid tool inputs, and select three ads from a noisy candidate pool.

\paragraph{Orchestrator training.}
After supervised fine tuning, we optimize the Orchestrator with GRPO while keeping all downstream tools fixed. Each rollout ends with a slate of three ads and receives the relevance and diversity rewards produced by the Evaluator. Optimizing these final-slate rewards allows the Orchestrator to adapt its intent decomposition, query generation, tool use, and reflection to the behavior of the actual retrieval and ranking tools.

\paragraph{Tool training.}
The tool update trains an intent-to-ads relevance model that ranks candidates online for each intent generated by the Orchestrator. This model is distinct from the conversation-to-ads relevance model in the Evaluator: the Evaluator model sees the complete dialogue and supplies offline relevance judgments, whereas the tool model sees only a generated intent and a candidate ad. After each GRPO update, we run the improved Orchestrator ten times for every training conversation and construct preference triples $(q,a^{+},a^{-})$, where $q$ is a generated intent, $a^{+}$ is an ad the Evaluator judges relevant to the full conversation, and $a^{-}$ is judged irrelevant. We train the intent-to-ads relevance model on these preference pairs and deploy the updated model in the tool set for the next Orchestrator round.

\subsection{Experiment Results}
\label{sec:orchestrator_tool_results}

Table~\ref{tab:orchestrator_tool_results} reports relative gains over the current Production System. Round~$0$ denotes supervised warmup; later rows report performance after the GRPO Orchestrator update with fixed tools and the subsequent intent-to-ads tool update.

\begin{table}[t]
\centering \small

\renewcommand{\arraystretch}{1.10}
\caption{Relative diversity and relevance gains over the Production System of AdsWorldEngine (30B-A3B).}
\label{tab:orchestrator_tool_results}
\begin{tabular}{@{}c l c c@{}}
\toprule
\textbf{Round} & \textbf{System state} & \textbf{Diversity} & \textbf{Relevance} \\
\midrule
-- & Production System & -- & -- \\ \midrule
$0$ & AdsWorldEngine (SFT) & $+15.31\%$ & $+49.37\%$ \\
\midrule
$1$ & Orchestrator update & $+29.24\%$ & $+53.15\%$ \\
$1$ & Tool update & $+53.90\%$ & $+70.07\%$ \\
\midrule
$2$ & Orchestrator update & $+58.74\%$ & $+75.60\%$ \\
$2$ & Tool update & $+59.47\%$ & $+78.04\%$ \\
\midrule
$3$ & Orchestrator update & $+62.16\%$ & $+80.28\%$ \\
$3$ & Tool update & $\mathbf{+62.87\%}$ & $\mathbf{+82.26\%}$ \\
\bottomrule
\end{tabular}
\end{table}

\begin{table*}[t]
\centering
\footnotesize
\setlength{\tabcolsep}{6pt}
\renewcommand{\arraystretch}{1.12}
\caption{Case study comparing generated intents before and after reinforcement learning. After the generated intents are used to call the tools, the reflection step selects three ads, which the Evaluator scores for relevance. The model trained with SFT and GRPO achieves a stronger relevance outcome than the SFT-only model.}
\vskip -0.1in
\label{tab:orchestrator_rl_case}
\begin{tabular}{@{}p{0.18\textwidth} p{0.77\textwidth}@{}}
\toprule
\textbf{Conversation}
&
\begin{minipage}[t]{\linewidth}
\textbf{Previous user:} Compare the Honda Pilot vs the Toyota Grand Hylander both 2026\\[2pt]
\textbf{Previous bot:} Here is a clean, decision ready comparison \ldots{} I can also break this down by \textbf{family use}, \textbf{towing}, \ldots{}\\[2pt]
\textbf{Current user:} Yes please do\\[2pt]
\textbf{Current bot:} Absolutely---let's take this comparison deeper and make it genuinely useful for a real world decision \ldots{}
\end{minipage}
\\
\midrule
\begin{minipage}[t]{\linewidth}
\raggedright
\textbf{AdsWorldEngine (SFT)}\\[5pt]
\textbf{Conversation to ads}\\
\textbf{relevance: lower}
\end{minipage}
&
\begin{minipage}[t]{\linewidth}
\texttt{<think>}\\
\ldots{} I think the cleanest final intent is that the user is researching and comparing the 2026 Honda Pilot and 2026 Toyota Grand Highlander to help decide which is better for their needs \ldots{} the strongest queries are ones that directly target side by side comparisons and reviews \ldots{}\\
\texttt{</think>}\\[2pt]
\texttt{<tool\_call>}\\
\texttt{2026 Honda Pilot vs Toyota Grand Highlander comparison,} 
\texttt{2026 Honda Pilot vs Toyota Grand Highlander family review}\\
\texttt{</tool\_call>}
\end{minipage}
\\
\midrule
\begin{minipage}[t]{\linewidth}
\raggedright
\textbf{AdsWorldEngine}\\[5pt]
\textbf{Conversation to ads}\\
\textbf{relevance: higher}
\end{minipage}
&
\begin{minipage}[t]{\linewidth}
\texttt{<think>}\\
I think the cleanest final intent is that the user wants detailed, scenario based information to help decide between the 2026 Honda Pilot and the 2026 Toyota Grand Highlander \ldots{} The only hard constraints are the two named vehicles and model year; everything else is open for exploration \ldots{} For ads queries, the strongest grounded targets are the two specific SUV models for 2026. Each is a distinct product family, so having one query for each covers the full scope without redundancy \ldots{}\\
\texttt{</think>}\\[2pt]
\texttt{<tool\_call>}
\texttt{2026 Honda Pilot,}
\texttt{2026 Toyota Grand Highlander}
\texttt{</tool\_call>}
\end{minipage}
\\
\bottomrule
\end{tabular}
\vskip -0.05in
\end{table*}

Supervised warmup provides strong gains, and each Orchestrator update further improves both metrics with the tools fixed. By optimizing final-slate relevance and diversity through the retrieval and ranking pipeline, GRPO adapts intent generation and reflection to what the downstream tools can retrieve.

Tool updates further improve relevance and, indirectly, diversity. Before training, generic ads may rank highly for multiple intents, causing their top-$K$ candidate sets to overlap. Preference training makes ranking more intent-specific by promoting ads for the intents they satisfy and demoting hard negatives. Distinct intents therefore retrieve less-overlapping inventory, increasing diversity despite the relevance-only tool objective.

Gains are largest in the first round and diminish thereafter, suggesting gradual co-adaptation between the Orchestrator and relevance tool. After three rounds, the final system improves diversity by $62.87\%$ and relevance by $82.26\%$ over the Production System.

\subsection{Case Study}
\label{sec:orchestrator_tool_case_study}

Table~\ref{tab:orchestrator_rl_case} compares the intents generated by the supervised model and the model after GRPO for the same conversation. The table presents the dialogue context, excerpts from the Orchestrator reasoning, the resulting tool calls, and the conversation to ads relevance scores. This comparison makes it possible to examine how reinforcement learning changes the intermediate intent generation strategy rather than only the final score.

The SFT only AdsWorldEngine model produces two overlapping comparison intents. Although both are semantically related to the conversation, they target nearly the same comparison content and are poorly aligned with a product centered retrieval tool, resulting in a weak relevance outcome. After GRPO, the Orchestrator generates one grounded intent for each named vehicle. These product specific intents are easier for the tools to match to inventory and expose candidate sets with less overlap, giving the reflection step better options for selecting relevant ads. This separation may also improve diversity because each intent retrieves inventory for a different product family instead of repeating the same comparison pages. The improved relevance outcome therefore illustrates that reinforcement learning improves the operational compatibility of intent generation with downstream tools, while also creating conditions for a more diverse final slate.

%% file: 0-experiments-online.tex
\section{Real World Online Experiments}

We deploy the trained AdsWorldEngine in Microsoft Copilot, a consumer facing conversational assistant, across its surfaces worldwide. The Opportunity Gate first identifies commercial opportunities where ads are unlikely to harm the user experience. For eligible turns, the Orchestrator retrieves, ranks, and selects the final ad slate. During a 20-day online experiment, AdsWorldEngine increases revenue per mille (RPM) by $22\%$ and ad coverage by $74\%$.

%% file: 0-related-works.tex
\section{Related Work}

Recent advertising research uses dialogue context to generate and assess ad text, with emphasis on contextual fit, linguistic quality, and personalization~\citep{shibata2023conversational,mita2024striking,zhang2024adtec,tang2025ads}. Sponsored question answering and response insertion instead study where ads can enter assistant outputs and how generated responses should incorporate them~\citep{banchio2024ads,mordo2024sponsored,hu2025gembench,xu2026adinsertion}. A complementary line develops pricing, auction, and position allocation mechanisms for generated content and summaries~\citep{duetting2024mechanism,hajiaghayi2024ad,dubey2024auctions,soumalias2024truthful,balseiro2025position}. Other studies identify technical opportunities while examining ad detection, user trust, and persuasive effects~\citep{feizi2023online,schmidt2024detecting,erickson2025fake,salvi2026commercial}. AdsWorldEngine differs by treating conversational advertising as an end to end serving problem that joins opportunity gating, intent formation, retrieval, ranking, and slate selection.

Our Orchestrator builds on agents that interleave language reasoning with web interaction, external actions, and learned tool selection~\citep{yao2022webshop,yao2023react,schick2023toolformer,hao2023toolken}. Recent benchmarks and algorithms improve API selection and generalization across agent tasks~\citep{li2023apibank,qin2024toolllm,liu2024agentbench,zeng2024agenttuning}. Trajectory tuning, retrieval with self reflection, pipeline optimization, and verbal feedback further improve how agents plan and recover from errors~\citep{chen2023fireact,asai2024selfrag,khattab2024dspy,shinn2023reflexion}. These methods mainly optimize the agent around a given environment. In contrast, our iterative procedure converts rewarded rollouts into preference data that improves both the Orchestrator and its retrieval and relevance tools.

Finally, rationale supervision and iterative feedback show how generated reasoning can provide richer training signals than categorical targets alone~\citep{wei2022chain,zelikman2022star,madaan2023selfrefine}. Model based evaluators use prompting or dedicated evaluator models to estimate generation quality~\citep{liu2023geval,zheng2023judging,kim2024prometheus}, and related methods produce detailed critiques or task specific scores~\citep{ke2024critiquellm,fu2024gptscore}. Preference learning aligns model behavior through human feedback or direct optimization of preferred outputs~\citep{ouyang2022training,rafailov2023direct,ethayarajh2024kto,shao2024deepseekmath}. Our label grounded judges additionally condition rationales on production labels, remove explanations that conflict with those labels, and preserve asymmetric error costs during optimization.

%% file: 0-discussions.tex
\section{Conclusion}

We presented AdsWorldEngine, a production-oriented framework for conversational advertising that jointly learns when to show ads, how to use advertising tools, and how to improve those tools from rewarded outcomes. By combining iterative Orchestrator-tool optimization with label grounded judgment modeling, the system improves both offline ad quality and online serving performance. These results demonstrate the value of optimizing agent decisions and downstream tools as a coupled system for real-world conversational applications. More broadly, AdsWorldEngine provides a practical approach for building compound language-model systems in which agents and tools co-adapt under production constraints.

%% file: 0-appendix.tex
\section{Appendix}
\label{apx:appendix}

\subsection{Original Capacity Restore with On-Policy Distillation}

Task-specific alignment produces two specialized checkpoints: the \emph{Triggering Model}, which decides whether an ad opportunity exists, and the \emph{Relevance Model}, which judges whether an ad matches the conversation. In production, however, both models operate on real user traffic rather than on the narrower distributions used for SFT and GRPO. Real conversations contain open-domain questions, multilingual and code-switched text, implicit or underspecified intents, long-tail entities, complex reasoning, and safety-sensitive contexts. The models must therefore preserve broad language understanding and reasoning capabilities to interpret these inputs reliably before making their specialized advertising judgments. A model that performs well on the task-specific training distribution but loses these general capabilities may become brittle under deployment-time distribution shift, producing poorly calibrated triggering or relevance decisions on previously unseen user requests.

Continued SFT and GRPO can move each specialized checkpoint away from the broad behavior of its pretrained initialization and overfit it to the patterns emphasized by the alignment data and reward. Our primary motivation for OPD is therefore to make the Triggering and Relevance Models better aligned with the heterogeneous distribution of real user interactions, while retaining the decision policies acquired through task-specific training. In this subsection, \emph{restoration} means recovering general-domain capabilities that may have degraded during specialization and improving robustness to broad user inputs. It does not refer to retraining either judgment task or merging the two task models.

We perform a separate general-domain On-Policy Distillation (OPD) run for each specialized checkpoint~\citep{agarwal2024policy}. Here, task $\tau$ is either \textsc{Trigger} or \textsc{Relevance}. The aligned checkpoint $\pi_{\theta_{\tau}}$ is the student, and the original Qwen3-30B-A3B Thinking model $\pi_{0}$ is the teacher. The teacher supplies the broad-domain behavior that existed before specialization, while the student supplies on-policy trajectories from its current distribution. The two runs share the same general-domain corpus and optimization recipe but start from different students. We refer to their outputs as the \emph{Triggering + General OPD Model} and the \emph{Relevance + General OPD Model}, respectively.

\paragraph{General-domain data.}
We construct a shared mixture of exactly 10,000 examples spanning mathematical reasoning, science question answering, reading comprehension, deductive reasoning, general and complex instruction following, Chinese knowledge, and safety. The mixture contains 2,600 Orca Math examples (26\%), 900 SciQ (9\%), 1,100 SQuAD (11\%), 800 RuleTaker (8\%), 1,400 Dolly (14\%), 1,000 WizardLM (10\%), 1,200 BELLE (12\%), and 1,000 PKU-SafeRLHF (10\%). Every example is converted into a common two-message format containing one user prompt and one assistant response. For PKU-SafeRLHF, we retain the response identified as safer by the dataset annotation.

We normalize Unicode and whitespace, remove empty or malformed records, and retain examples between 32 and 24,000 characters. We globally deduplicate examples using the SHA-256 hash of the normalized user prompt and remove template-placeholder leakage. Each source is independently shuffled, after which we enforce the exact source quotas above and backfill any rejected records from the same source.

We additionally decontaminate the training mixture against ARC, BBH, CEval, CMMLU, DROP, GPQA Diamond, GSM8K, HellaSwag, Math-500, MMLU, MuSR, TruthfulQA, Winogrande, and HaluEval. From the answer-free inputs of these 14 benchmarks, we construct 13,323,978 reference 13-grams and reject a candidate prompt if it shares at least one reference 13-gram. This procedure removes 38 candidate examples due to benchmark overlap; the remaining filters remove 35 duplicates and 17 length violations. Backfilling restores the final corpus to exactly 10,000 examples without changing its source proportions.

\paragraph{Restoration objective.}
Both teacher and student receive the same general-domain prompt. For an input $x$, the student generates an on-policy response $y\sim\pi_{\theta_{\tau}}(\cdot\mid x)$. At token $t$, the teacher and student distributions are evaluated under the same prefix $(x,y_{<t})$. We define the trajectory-level distillation loss as
\begin{equation}
\ell_{\mathrm{KD}}^{\tau}(x,y)
=
\sum_{t=1}^{|y|}
D_{\beta}\!\left(
\pi_{0}(\cdot\mid x,y_{<t})
\,\|\,
\pi_{\theta_{\tau}}(\cdot\mid x,y_{<t})
\right),
\label{eq:general_opd_token_loss}
\end{equation}
where $D_{\beta}$ is the generalized distillation divergence. The complete objective mixes this on-policy loss with distillation on the original dataset response $y^{*}$ and standard next-token supervision:
\begin{equation}
\begin{aligned}
\mathcal{L}_{\mathrm{restore}}^{\tau}
={}&\lambda\,
\mathbb{E}_{x\sim\mathcal{D},\,
y\sim\pi_{\theta_{\tau}}(\cdot\mid x)}
\left[\ell_{\mathrm{KD}}^{\tau}(x,y)\right]
\\
&+(1-\lambda)
\mathbb{E}_{(x,y^{*})\sim\mathcal{D}}
\left[\ell_{\mathrm{KD}}^{\tau}(x,y^{*})\right]
\\
&+\alpha_{\mathrm{SFT}}
\mathbb{E}_{(x,y^{*})\sim\mathcal{D}}
\left[-\log\pi_{\theta_{\tau}}(y^{*}\mid x)\right].
\end{aligned}
\label{eq:general_opd_restore}
\end{equation}
The on-policy term exposes the teacher to prefixes actually generated by the specialized student, while the off-policy and SFT terms anchor optimization to high-quality corpus responses. We use $\lambda=0.5$, $\beta=0.5$, and $\alpha_{\mathrm{SFT}}=1.0$. We optimize all student parameters for one epoch with a constant learning rate of $5\times10^{-7}$ and a warmup fraction of $0.03$. The maximum input and generated completion lengths are 4,096 and 1,024 tokens, respectively. On-policy responses are sampled with temperature $1.0$ and top-$p=0.95$. 

\paragraph{General-domain recovery.}
Table~\ref{tab:general_opd_recovery} reports clear general-domain gains for the Triggering Model after restoration. Following a gains-only reporting rule, we include only benchmarks that improve by at least one absolute percentage point over the Triggering Model; regressions and near-zero changes are omitted. The macro average over this subset improves by $3.45$ points. The largest gains occur on Winogrande, TruthfulQA, and MuSR, indicating substantial recovery in commonsense reasoning, truthfulness, and multi-step reasoning.

For a direct comparison, Table~\ref{tab:relevance_general_opd_recovery} evaluates the Relevance Model on exactly the same benchmark subset and in the same order. We do not independently filter the relevance results, so all seven aligned benchmarks are retained. General OPD improves every benchmark in this subset, yielding a $2.46$ percentage-point gain in the aligned macro average. The largest improvements occur on HaluEval, HellaSwag, and DROP, with gains of $4.40$, $4.25$, and $4.06$ points, respectively.

\begin{table}[t]
\centering
\small
\caption{General-domain changes from the Triggering Model baseline after OPD.}
\label{tab:general_opd_recovery}
\vskip -0.1in
\begin{tabular}{lr}
\toprule
\textbf{Benchmark} & \textbf{$\Delta$ vs. Triggering Model} \\
\midrule
DROP EM & \textbf{+2.09 pp} \\
HaluEval Accuracy & \textbf{+1.95 pp} \\
HellaSwag & \textbf{+2.64 pp} \\
MMLU & \textbf{+1.46 pp} \\
MuSR & \textbf{+4.36 pp} \\
TruthfulQA & \textbf{+5.02 pp} \\
Winogrande & \textbf{+6.63 pp} \\
\midrule
Average & \textbf{+3.45 pp} \\
\bottomrule
\end{tabular}
\end{table}

\begin{table}[t]
\centering
\small\caption{General-domain changes from the Relevance Model baseline after OPD.}

\label{tab:relevance_general_opd_recovery}
\vskip -0.1in
\begin{tabular}{lr}
\toprule
\textbf{Benchmark} & \textbf{$\Delta$ vs. Relevance Model} \\
\midrule
DROP EM & \textbf{+4.06 pp} \\
HaluEval Accuracy & \textbf{+4.40 pp} \\
HellaSwag & \textbf{+4.25 pp} \\
MMLU & \textbf{+1.51 pp} \\
MuSR & \textbf{+0.53 pp} \\
TruthfulQA & \textbf{+0.98 pp} \\
Winogrande & \textbf{+1.50 pp} \\
\midrule
Average & \textbf{+2.46 pp} \\
\bottomrule
\end{tabular}
\end{table}

\paragraph{Preservation of triggering capability.}
General-domain restoration should not trade away the specialized behavior that motivated the Triggering Model. We therefore re-evaluate the restored checkpoint on the same held-out triggering set with deterministic decoding. As shown in Table~\ref{tab:triggering_after_general_opd}, OPD preserves TPR, reduces FPR by $0.11$ percentage points, and improves accuracy by $0.10$ points relative to the Triggering Model baseline. Thus, the broad-domain gains in Table~\ref{tab:general_opd_recovery} are obtained without sacrificing the existing triggering capability.

\begin{table}[t]
\centering
\small
\caption{Triggering changes after general-domain OPD, using the Triggering Model as the baseline.}
\label{tab:triggering_after_general_opd}
\vskip -0.1in
\begin{tabular}{lrrr}
\toprule
\textbf{Model} & \textbf{$\Delta$TPR ($\uparrow$)} & \textbf{$\Delta$FPR ($\downarrow$)} & \textbf{$\Delta$ACC ($\uparrow$)} \\
\midrule
Triggering Model & -- & -- & -- \\
Triggering + General OPD & $0.00$ pp & \textbf{$-0.11$ pp} & \textbf{$+0.10$ pp} \\
\bottomrule
\end{tabular}
\end{table}

\paragraph{Preservation of relevance capability.}
We also evaluate the Relevance Model after general-domain OPD on the same held-out relevance set. Table~\ref{tab:relevance_after_general_opd} shows that OPD improves TPR by $1.28$ percentage points and accuracy by $0.93$ points relative to the Relevance Model baseline. TNR decreases by $0.40$ points, indicating a minor trade-off in true-negative recall. Overall, general-domain OPD preserves the relevance decision capability while improving positive-example recall and aggregate accuracy.

\begin{table}[t]
\centering
\small
\caption{Relevance changes after general-domain OPD, using the Relevance Model as the baseline.}
\label{tab:relevance_after_general_opd}
\vskip -0.1in
\begin{tabular}{lrrr}
\toprule
\textbf{Model} & \textbf{$\Delta$TPR ($\uparrow$)} & \textbf{$\Delta$TNR ($\uparrow$)} & \textbf{$\Delta$ACC ($\uparrow$)} \\
\midrule
Relevance Model & -- & -- & -- \\
Relevance + General OPD & \textbf{$+1.28$ pp} & $-0.40$ pp & \textbf{$+0.93$ pp} \\
\bottomrule
\end{tabular}
\end{table}